\documentclass[11pt]{revtex4}    
\usepackage{amssymb,epsf}
\usepackage{latexsym}
\begin{document}

\title{Rotating Black Branes in Brans-Dicke-Born-Infeld Theory}
\author{S. H. Hendi\footnote{hendi@mail.yu.ac.ir}}

\address{Physics Department,
College of Sciences, Yasouj University, Yasouj
75914, Iran\\
 Research Institute for Astrophysics and Astronomy of Maragha
(RIAAM), P.O. Box 55134-441, Maragha, Iran}

\begin{abstract}
In this paper, we present a new class of charged rotating black brane
solutions in the higher dimensional Brans-Dicke-Born-Infeld theory and
investigate their properties. Solving the field equations directly is a
non-trivial task because they include the second derivatives of the scalar
field. We remove this difficulty through a conformal transformation. Also,
we find that the suitable Lagrangian of Einstein-Born-Infeld-dilaton gravity
is not the same as presented in \cite{DHSR}. We show that the given
solutions can present black brane, with inner and outer event horizons, an
extreme black brane or a naked singularity provided the parameters of the
solutions are chosen suitably. These black brane solutions are neither
asymptotically flat nor (anti)-de Sitter. Then we calculate finite Euclidean
action, the conserved and thermodynamic quantities through the use of
counterterm method. Finally, we argue that these quantities satisfy the
first law of thermodynamics, and the entropy does not follow the area law.
\end{abstract}

\maketitle

\address{ Physics Department,
College of Sciences, Yasouj University, Yasouj 75914, Iran;\\
 Research Institute for Astrophysics and Astronomy of Maragha
(RIAAM), P.O. Box 55134-441, Maragha, Iran}

\section{Introduction}

Lately there have been some renewed interest in the Brans-Dicke (BD) theory
of gravitation \cite{JBD}. On one hand, it is important for cosmological
inflation models \cite{LaSte}, in which the scalar field allows the
inflationary epoch to end via bubble nucleation without the need for
fine-tuning cosmological parameters (the "graceful exit" problem). Also, it
was found that in the low-energy regime, the theory of fundamental strings
can be reduced to an effective BD one \cite{LavFr}.

Because scalar-tensor gravitation can agree with general relativity (GR) in
the post-Newtonian limit, it is important to study strong field examples in
which the two theories may give different predictions. These examples may
not only provide further experimental and observational tests that might
distinguish between GR and scalar-tensor gravitation, but they may also
illuminate the structure of both theories.

The BD theory incorporates the Mach principle, which states that the
phenomenon of inertia must arise from accelerations with respect to the
general mass distribution of the universe. This theory is self-consistent,
complete and for $|\omega |\geq 500$ in accord with solar system
observations and experiments \cite{Will}, where $\omega $ is an adjustable
parameter. In this theory, the matter couples minimally to the metric and
not directly to scalar field. Indeed, the scalar field does not exert any
direct influence on matter, its only role is that of participating in the
field equations that determine the geometry of the spacetime. More recently,
many authors have investigated the gravitational collapse and black hole
formation in the BD theory \cite{AllBD,DPH}.

Till now, nonlinear charged rotating black hole solutions for an arbitrary
value of $\omega $ has not been constructed. In this paper, we want to
construct exact rotating black brane solutions in BD-Born-Infeld (BDBI)
theory for an arbitrary value of $\omega $ and investigate their properties.
One can find that solving the field equations directly is a non-trivial
task, because they include the second derivatives of the scalar field. We
remove this difficulty through a conformal transformation. By using this
transformation, the BDBI action reduce to Einstein-Born-Infeld-dilaton
(EBId) action, and one can solve their field equations analytically.

The idea of the non-linear electrodynamics (BI) was first introduced in $%
1934 $ by Born and Infeld in order to obtain a finite value for the
self-energy of point-like charges \cite{BI}. Although it become less popular
with the introduction of QED, in recent years, the BI action has been
occurring repeatedly with the development of superstring theory, where the
dynamics of $D$-branes is governed by the BI action \cite{Frad,Call}.
Lately, black hole solutions in BI gravity with or without a cosmological
constant have been considered by many authors \cite{BIpaper1,BIpaper2}. Both
of the Lagrangians of EBId gravity presented here and in \cite{DHSR} show
similar asymptotic behavior but only the one we considered is consistent
with BD theory.

The outline of this paper is organized as follows. In Sec. \ref{Con}, we
give a brief review of the field equations of BDBI theory in Jordan (or
string) and Einstein frames. In Sec. \ref{Sol}, we obtain charge rotating
solution in $(n+1)$-dimensions with $k$ rotation parameters and investigate
their (asymptotic) properties. Sec. \ref{Therm} is devote to calculation of
the finite action, the conserved and thermodynamic quantities of the $(n+1)$%
-dimensional black brane solutions with a complete set of rotational
parameters. We finish our paper with some concluding remarks.

\bigskip

\section{Field Equation and Conformal Transformation\label{Con}}

In $n+1$ dimensions, the action of the BDBI theory with one scalar field $%
\Phi $ and a self-interacting potential $V(\Phi )$ can be written as
\begin{equation}
I_{G}=-\frac{1}{16\pi }\int_{\mathcal{M}}d^{n+1}x\sqrt{-g}\left( \Phi
\mathcal{R}\text{ }-\frac{\omega }{\Phi }(\nabla \Phi )^{2}-V(\Phi
)+L(F)\right) ,  \label{I1}
\end{equation}%
where $\mathcal{R}$ is the Ricci scalar, $\omega $ is the coupling constant,
$\Phi $ denotes the BD scalar field and $V(\Phi )$ is a self-interacting
potential for $\Phi $ and $L(F)$ is the Lagrangian of BI
\begin{equation}
L(F)=4\beta ^{2}\left( 1-\sqrt{1+\frac{F^{2}}{2\beta ^{2}}}\right) ,
\label{ActEB}
\end{equation}%
In Eq. (\ref{ActEB}), the constant $\beta $\ is called BI parameter with
dimension of mass, $F^{2}=F^{\mu \nu }F_{\mu \nu }$\ where $F_{\mu \nu
}=\partial _{\mu }A_{\nu }-\partial _{\nu }A_{\mu }$\ is electromagnetic
tensor field and $A_{\mu }$\ is the vector potential. In the limit $\beta
\rightarrow \infty $, $L(F)$ reduces to the standard Maxwell form $%
L(F)=-F^{2}$, while $L(F)\rightarrow 0$ as $\beta \rightarrow 0$. Varying
the action (\ref{I1}) with respect to the metric, scalar and vector fields
give the field equations as
\begin{eqnarray}
&&G_{\mu \nu }=\frac{\omega }{\Phi ^{2}}\left( \nabla _{\mu }\Phi \nabla
_{\nu }\Phi -\frac{1}{2}g_{\mu \nu }(\nabla \Phi )^{2}\right) -\frac{V(\Phi )%
}{2\Phi }g_{\mu \nu }+\frac{1}{\Phi }\left( \nabla _{\mu }\nabla _{\nu }\Phi
-g_{\mu \nu }\nabla ^{2}\Phi \right)  \nonumber \\
&&\hspace{1cm}+\frac{2}{\Phi }\left( \frac{1}{4}g_{\mu \nu }L(F)+\frac{%
F_{\mu \lambda }F_{\phantom{\lambda}{\nu}}^{\lambda }}{\sqrt{1+\frac{F^{2}}{%
2\beta ^{2}}}}\right) ,  \label{fil1} \\
&&\nabla ^{2}\Phi =\frac{1}{2\left[ \left( n-1\right) \omega +n\right] }\Big[%
\left( (n+1)L(F)+\frac{4F^{2}}{\sqrt{1+\frac{F^{2}}{2\beta ^{2}}}}\right)
\nonumber \\
&&+\left( (n-1)\Phi \frac{dV(\Phi )}{d\Phi }-\left( n+1\right) V(\Phi
)\right) \Big],  \label{fil2} \\
&&\partial _{\mu }\left( \frac{\sqrt{-g}F^{\mu \nu }}{\sqrt{1+\frac{F^{2}}{%
2\beta ^{2}}}}\right) =0,  \label{fil3}
\end{eqnarray}%
where $G_{\mu \nu }$ and $\nabla _{\mu }$ are the Einstein tensor and
covariant differentiation corresponding to the metric $g_{\mu \nu }$
respectively. Solving the field equations (\ref{fil1})-(\ref{fil3}) directly
is a non-trivial task because the right hand side of Eq. (\ref{fil1})
includes the second derivatives of the scalar. We can remove this difficulty
by the conformal transformation
\begin{eqnarray}
\bar{g}_{\mu \nu } &=&\Phi ^{2/(n-1)}g_{\mu \nu },  \nonumber \\
\bar{\Phi} &=&\frac{n-3}{4\alpha }\ln \Phi ,  \label{conf}
\end{eqnarray}
where
\begin{equation}
\alpha =(n-3)/\sqrt{4(n-1)\omega +4n}  \label{alpha}
\end{equation}
One may note that $\alpha$ goes to zero as $\omega$ goes to infinity and the
BD theory reduces to Einstein theory. By this transformation, the action (%
\ref{I1}) transforms to
\begin{equation}
\bar{I}_{G}=-\frac{1}{16\pi }\int_{\mathcal{M}}d^{n+1}x\sqrt{-\bar{g}}%
\left\{ \bar{\mathcal{R}}\text{ }-\frac{4}{n-1}(\overline{\nabla }\bar{\Phi}%
)^{2}-\bar{V}(\bar{\Phi})+\bar{L}(\bar{F},\bar{\Phi})\right\} ,  \label{I2}
\end{equation}%
where $\bar{\mathcal{R}}$ and $\ \bar{\nabla}$ are the Ricci scalar and
covariant differentiation corresponding to the metric $\bar{g}_{\mu \nu }$,
and $\bar{V}(\bar{\Phi})$ is
\[
\bar{V}(\bar{\Phi})=\Phi ^{-(n+1)/(n-1)}V(\Phi )
\]%
The Born-Infeld Lagrangian coupled to a dilaton field, $\bar{L}(\bar{F},\bar{%
\Phi})$ corresponding to the metric $\bar{g}_{\mu \nu }$ is given by
\begin{equation}
\bar{L}(\bar{F},\bar{\Phi})=4\beta ^{2}e^{-4\alpha (n+1)\bar{\Phi}%
/[(n-1)(n-3)]}\left( 1-\sqrt{1+\frac{e^{16\alpha \bar{\Phi}/[(n-1)(n-3)]}%
\bar{F}^{2}}{2\beta ^{2}}}\right) ,  \label{LFP}
\end{equation}%
In the limit $\beta \rightarrow \infty $, $\bar{L}(\bar{F},\bar{\Phi})$
reduces to the standard Maxwell field coupled to a dilaton field
\begin{equation}
\bar{L}(\bar{F},\bar{\Phi})=-e^{-4\alpha \bar{\Phi}/(n-1)}\bar{F}^{2}.
\end{equation}%
On the other hand, $\bar{L}(\bar{F},\bar{\Phi})\rightarrow 0$ as $\beta
\rightarrow 0$. It is convenient to set
\begin{equation}
\bar{L}(\bar{F},\bar{\Phi})=4\beta ^{2}e^{-4\alpha (n+1)\bar{\Phi}%
/[(n-1)(n-3)]}{\bar{L}}(\bar{Y}),
\end{equation}%
where
\begin{eqnarray}
{\bar{L}}(\bar{Y}) &=&1-\sqrt{1+\bar{Y}},  \label{LY} \\
\bar{Y} &=&\frac{e^{16\alpha \bar{\Phi}/[(n-1)(n-3)]}\bar{F}^{2}}{2\beta ^{2}%
}.  \label{Y}
\end{eqnarray}%
It is notable that this action is different from the action of EBId that has
been presented in \cite{DHSR}. Due to the fact that the action of EBId in
\cite{DHSR} is not consistent with conformal transformation, one can find
that Eq. (\ref{I2}) is the suitable action for EBId gravity.

Varying the action (\ref{I2}) with respect to $\bar{g}_{\mu \nu }$, $\bar{%
\Phi}$ and $\bar{F}_{\mu \nu }$, we obtain equations of motion as
\begin{eqnarray}
&&\bar{\mathcal{R}}_{\mu \nu }=\frac{4}{n-1}\left( \bar{\nabla}_{\mu }\bar{%
\Phi}\bar{\nabla}_{\nu }\bar{\Phi}+\frac{1}{4}\bar{V}\bar{g}_{\mu \nu
}\right) -\frac{1}{(n-1)}\bar{L}(\bar{F},\bar{\Phi})\bar{g}_{\mu \nu }
\nonumber \\
&&-2e^{-4\alpha \bar{\Phi}/(n-1)}\partial _{\bar{Y}}{\bar{L}}(\bar{Y})\left(
\bar{F}_{\mu \lambda }\bar{F}_{\nu \phantom{\nu}{\lambda}}^{\lambda }-\frac{2%
\bar{F}^{2}}{(n-1)}\bar{g}_{\mu \nu }\right) ,  \label{fild1} \\
&&\bar{\nabla}^{2}\bar{\Phi}=\frac{n-1}{8}\frac{\partial \bar{V}}{\partial
\bar{\Phi}}+\frac{\alpha }{2(n-3)}\left( (n+1)\bar{L}(\bar{F},\bar{\Phi}%
)-8e^{-4\alpha \bar{\Phi}/(n-1)}\partial _{\bar{Y}}{\bar{L}}(\bar{Y})\bar{F}%
^{2}\right) ,  \label{fild2} \\
&&\partial _{\mu }\left[ \sqrt{-\bar{g}}e^{-4\alpha \bar{\Phi}%
/(n-1)}\partial _{\bar{Y}}{\bar{L}}(\bar{Y})\bar{F}^{\mu \nu }\right] =0.
\label{fild3}
\end{eqnarray}%
Therefore, if $(\bar{g}_{\mu \nu },\bar{F}_{\mu \nu },\bar{\Phi})$ is the
solution of Eqs. (\ref{fild1})-(\ref{fild3}) with potential $\bar{V}(\bar{%
\Phi})$, then
\begin{equation}
\left[ g_{\mu \nu },F_{\mu \nu },\Phi \right] =\left[ \exp \left( -\frac{%
8\alpha \bar{\Phi}}{\left( n-1\right) (n-3)}\right) \bar{g}_{\mu \nu },\bar{F%
}_{\mu \nu },\exp \left( \frac{4\alpha \bar{\Phi}}{n-3}\right) \right]
\label{trans}
\end{equation}%
is the solution of Eqs. (\ref{fil1})-(\ref{fil3}) with potential $V(\Phi )$.

\section{Charged Rotating Solutions In $n+1$ dimensions with $k$ Rotation
Parameters\label{Sol}}

Here we construct the $(n+1)$-dimensional\ solutions of BD theory with $%
n\geq 4$ and the quadratic potential
\begin{eqnarray*}
V(\Phi )=2\Lambda \Phi ^{2}
\end{eqnarray*}%
Applying the conformal transformation (\ref{conf}), the potential $\bar{V}(%
\bar{\Phi})$ becomes
\begin{equation}
\bar{V}(\bar{\Phi})=2\Lambda \exp \left( \frac{4\alpha \bar{\Phi}}{n-1}%
\right) ,  \label{Liovpot}
\end{equation}%
which is a Liouville-type potential. Thus, the problem of solving Eqs. (\ref%
{fil1})-(\ref{fil3}) with quadratic potential reduces to the problem of
solving Eqs. (\ref{fild1})-(\ref{fild3}) with Liouville-type potential.

The rotation group in $n+1$ dimensions is $SO(n)$ and therefore the number
of independent rotation parameters for a localized object is equal to the
number of Casimir operators, which is $[n/2]\equiv k$, where $[x]$ is the
integer part of $x$. The solutions of the field equations (\ref{fild1})-(\ref%
{fild3}) with $k$ rotation parameter $a_{i}$, and Liouville-type potential
is \cite{SDRP}
\begin{eqnarray}
d\bar{s}^{2} &=&-f(r)\left( \Xi dt-{{\sum_{i=1}^{k}}}a_{i}d\varphi
_{i}\right) ^{2}+\frac{r^{2}}{l^{4}}R^{2}(r){{\sum_{i=1}^{k}}}\left(
a_{i}dt-\Xi l^{2}d\varphi _{i}\right) ^{2}  \nonumber \\
&&-\frac{r^{2}}{l^{2}}R^{2}(r){\sum_{i<j}^{k}}(a_{i}d\varphi
_{j}-a_{j}d\varphi_{i})^{2}+\frac{dr^{2}}{f(r)}+\frac{r^{2}}{l^{2}}
R^{2}(r)dX^{2},  \nonumber \\
\Xi ^{2} &=&1+\sum_{i=1}^{k}\frac{a_{i}^{2}}{l^{2}},  \nonumber \\
\bar{F}_{tr} &=&\frac{q\Xi \beta e^{4\alpha \bar{\Phi}/(n-1)}}{\sqrt{%
q^{2}e^{8\alpha \bar{\Phi}/(n-3)}+\beta ^{2}r^{2n-2}R^{2n-2}}},\text{ \ \ \
\ \ \ \ \ }\bar{F}_{\varphi_{i} r}=-\frac{a_{i}}{\Xi }\bar{F}_{tr}.
\label{dil-metric}
\end{eqnarray}%
where $l$ is a constant, called length scale and $dX^{2}$ is the
flat Euclidean metric on $(n-k-1)$-dimensional
submanifold with volume $\omega _{n-k-1}$. Here $f(r)$, $R(r)$ and $\bar{\Phi%
}(r)$ are
\begin{eqnarray}
f(r) &=&\left( \frac{(1+\alpha ^{2})^{2}r^{2}}{(n-1)}\right) \left( \frac{%
2\Lambda \left( \frac{r}{c}\right) ^{-2\gamma }}{(\alpha ^{2}-n)}+\frac{%
4(n-3)\beta ^{2}\left( \frac{r}{c}\right) ^{2\gamma (n+1)/(n-3)}}{\lambda }%
\right) -  \nonumber \\
&&\frac{m}{r^{(n-1)(1-\gamma )-1}}-\frac{4(1+\alpha ^{2})^{2}q^{2}\left(
\frac{r}{c}\right) ^{2\gamma (n-2)}}{\lambda r^{2(n-2)}}\digamma (\eta ),
\label{Fr} \\
\digamma (\eta ) &=&\frac{(n-3)\sqrt{1+\eta }}{(n-1)\eta }-\frac{1}{\Upsilon
}\text{{\ }}_{2}F_{1}\left( \left[ \frac{1}{2},\frac{(n-3)\Upsilon }{2(n-1)}%
\right] ,\left[ 1+\frac{(n-3)\Upsilon }{2(n-1)}\right] ,-\eta \right) ,
\nonumber \\
\eta &=&\frac{q^{2}\left( \frac{r}{c}\right) ^{2\gamma (n-1)(n-5)/(n-3)}}{%
\beta ^{2}r^{2(n-1)}},  \nonumber \\
\Upsilon &=&\frac{\alpha ^{2}+n-2}{2\alpha ^{2}+n-3}  \nonumber \\
\lambda &=&(3n-1)\alpha ^{2}+n(n-3), \\
R(r) &=&\exp (\frac{2\alpha \bar{\Phi}}{n-1})=\left( \frac{r}{c}\right)
^{-\gamma },  \label{Rr} \\
\bar{\Phi}(r) &=&-\frac{(n-1)\gamma }{2\alpha }\ln \left( \frac{r}{c}\right)
,  \label{Phir}
\end{eqnarray}%
where $c$ is an arbitrary constant and $\gamma =\alpha ^{2}/(\alpha ^{2}+1)$%
. Using the conformal transformation (\ref{trans}), the ($n+1$)-dimensional
rotating solutions of BD theory with $k$ rotation parameters can be obtained
as
\begin{eqnarray}
ds^{2} &=&-U(r)\left( \Xi dt-{{\sum_{i=1}^{k}}}a_{i}d\varphi _{i}\right)
^{2}+\frac{r^{2}}{l^{4}}H^{2}(r){{\sum_{i=1}^{k}}}\left( a_{i}dt-\Xi
l^{2}d\varphi _{i}\right) ^{2}  \nonumber \\
&&-\frac{r^{2}}{l^{2}}H^{2}(r){\sum_{i<j}^{k}}(a_{i}d\varphi
_{j}-a_{j}d\varphi _{i})^{2}+\frac{dr^{2}}{V(r)}+\frac{r^{2}}{l^{2}}%
H^{2}(r)dX^{2},  \label{Met3}
\end{eqnarray}%
where $U(r)$, $V(r)$, $H(r)$ and $\Phi (r)$ are
\begin{eqnarray}
U(r) &=&\left( \frac{r}{c}\right) ^{4\gamma /(n-3)}f(r),  \label{Ur} \\
V(r) &=&\left( \frac{r}{c}\right) ^{-4\gamma /(n-3)}f(r),  \label{Vr} \\
H(r) &=&\left( \frac{r}{c}\right) ^{-\gamma (n-5)/(n-3)},  \label{Hr} \\
\Phi (r) &=&\left( \frac{r}{c}\right) ^{-2\gamma (n-1)/(n-3)}.  \label{Pr}
\end{eqnarray}%
The electromagnetic field becomes:
\begin{equation}
F_{tr}=\frac{q\Xi \beta \left( \frac{c}{r}\right) ^{-4\gamma /(n-3)}}{\sqrt{%
q^{2}+\beta ^{2}r^{2(n-1)[1-\gamma (n-5)/(n-3)]}c^{2\gamma (n-1)(n-5)/(n-3)}}%
},\text{\ \ \ }F_{\varphi_{i} r}=-\frac{a_{i}}{\Xi }F_{tr}.  \label{Ftr}
\end{equation}%
It is worth to note that the scalar field $\Phi (r)$ and electromagnetic
field $F_{\mu \nu }$ become zero as $r$ goes to infinity. These solutions
reduce to the solutions presented in Ref. \cite{DPH} as $\beta$ goes to
infinity. In the absence of a nontrivial dilaton ($\alpha =\gamma =0$ or $%
\omega \rightarrow \infty$), the above solutions reduce to those of Ref.
\cite{Rastegar} and in the limit $\beta \rightarrow \infty $ and $\omega
\rightarrow \infty $, these solutions reduce to the solutions of Refs. \cite%
{Lem,Deh3}. It is also notable to mention that these solutions are valid for
all values of $\omega $.

\subsection{Properties of the solutions}

In order to study the general structure of these solutions, we first look
for the essential singularities. One can show that the Kretschmann scalar $%
R_{\mu \nu \lambda \kappa }R^{\mu \nu \lambda \kappa }$ diverges at $r=0$,
and therefore there is a curvature singularity located at $r=0$. Seeking
possible black hole solutions, we turn to look for the existence of
horizons. Because of the presence of the hypergeometric function in the
equation $f(r)=0$, the radius of the event horizon cannot be found
explicitly. The roots of the metric function $f(r)$ are located at%
\begin{eqnarray}
&&-\frac{r_{+}^{\gamma (n-1)-n}m}{4(1+\alpha ^{2})^{2}}+\frac{\left( \frac{%
r_{+}}{c}\right) ^{-2\gamma }\Lambda }{2(n-1)(\alpha ^{2}-n)}+\frac{%
(n-3)\beta ^{2}\left( \frac{r_{+}}{c}\right) ^{2\gamma (n+1)/(n-3)}(1-\sqrt{%
1+\eta _{+}})}{(n-1)\lambda }  \nonumber \\
&&+\frac{r_{+}^{2[\gamma (n-2)-(n-1)]}q^{2}\text{{\ }}_{2}F_{1}\left( \left[
\frac{1}{2},\frac{(n-3)\Upsilon }{2(n-1)}\right] ,\left[ 1+\frac{%
(n-3)\Upsilon }{2(n-1)}\right] ,-\eta _{+}\right) }{\lambda c^{2\gamma
(n-2)}\Upsilon }=0
\end{eqnarray}
The angular velocities $\Omega _{i}$ are \cite{dehbordshah}
\begin{equation}
\Omega _{i}=\frac{a_{i}}{\Xi l^{2}},  \label{angvel}
\end{equation}%
and the temperature may be obtained through the use of definition of surface
gravity, $\kappa $,
\begin{equation}
T_{+}=\frac{1}{\mathcal{\beta }_{+}}=\frac{\kappa }{2\pi }=\frac{1}{2\pi }%
\sqrt{-\frac{1}{2}\left( \nabla _{\mu }\chi _{\nu }\right) \left( \nabla
^{\mu }\chi ^{\nu }\right) },
\end{equation}%
where $\chi $ is the Killing vector given by%
\begin{equation}
\chi =\partial _{t}+{{{\sum_{i=1}^{k}}}}\Omega _{i}\partial _{\phi _{i}}.
\label{Kil}
\end{equation}
One obtains
\begin{eqnarray}
T_{+} &=&-\frac{(1+\alpha ^{2})r_{+}}{2\pi \Xi }\Big[\frac{(\alpha
^{2}-n)r_{+}^{\gamma (n-1)}m}{2(1+\alpha ^{2})^{2}r^{n}}-\frac{8\alpha
^{2}\beta ^{2}\left( \frac{r_{+}}{c}\right) ^{2\gamma (n+1)/(n-3)}}{\lambda }%
\left( 1-\sqrt{1+\eta _{+}}\right)  \nonumber \\
&&-\frac{2(\alpha ^{2}-n)\left( \frac{r_{+}}{c}\right) ^{2\gamma (n-2)}q^{2}%
}{\lambda \Upsilon r_{+}^{2(n-1)}}\text{{\ }}_{2}F_{1}\left( \left[ \frac{1}{%
2},\frac{(n-3)\Upsilon }{2(n-1)}\right] ,\left[ 1+\frac{(n-3)\Upsilon }{%
2(n-1)}\right] ,-\eta _{+}\right) \Big]  \nonumber \\
&=&-\frac{(1+\alpha ^{2})r_{+}}{2\pi \Xi (n-1)}\left[ \left( \frac{r_{+}}{c}%
\right) ^{-2\gamma }\Lambda -2\beta ^{2}\left( \frac{r_{+}}{c}\right)
^{2\gamma (n+1)/(n-3)}\left( 1-\sqrt{1+\eta _{+}}\right) \right]  \label{Tem}
\end{eqnarray}%
which shows that the temperature of the solution is invariant under the
conformal transformation (\ref{conf}). This is due to the fact that the
conformal parameter is regular at the horizon.

{\Large Asymptotic Behavior:}

$\alpha ^{2}=n$: The solution is ill-defined for $\alpha ^{2}=n$ with a
quadratic potential ($\Lambda \neq 0$).

$\alpha ^{2}>n$: In this case, as $r$ goes to infinity the dominant term in
Eq. (\ref{Ur}) is the second term ($m$ term), and therefore the spacetime
has a cosmological horizon for positive values of the mass parameter,
despite the sign of the cosmological constant $\Lambda $.

\begin{figure}[tbp]
\epsfxsize=5cm \centerline{\epsffile{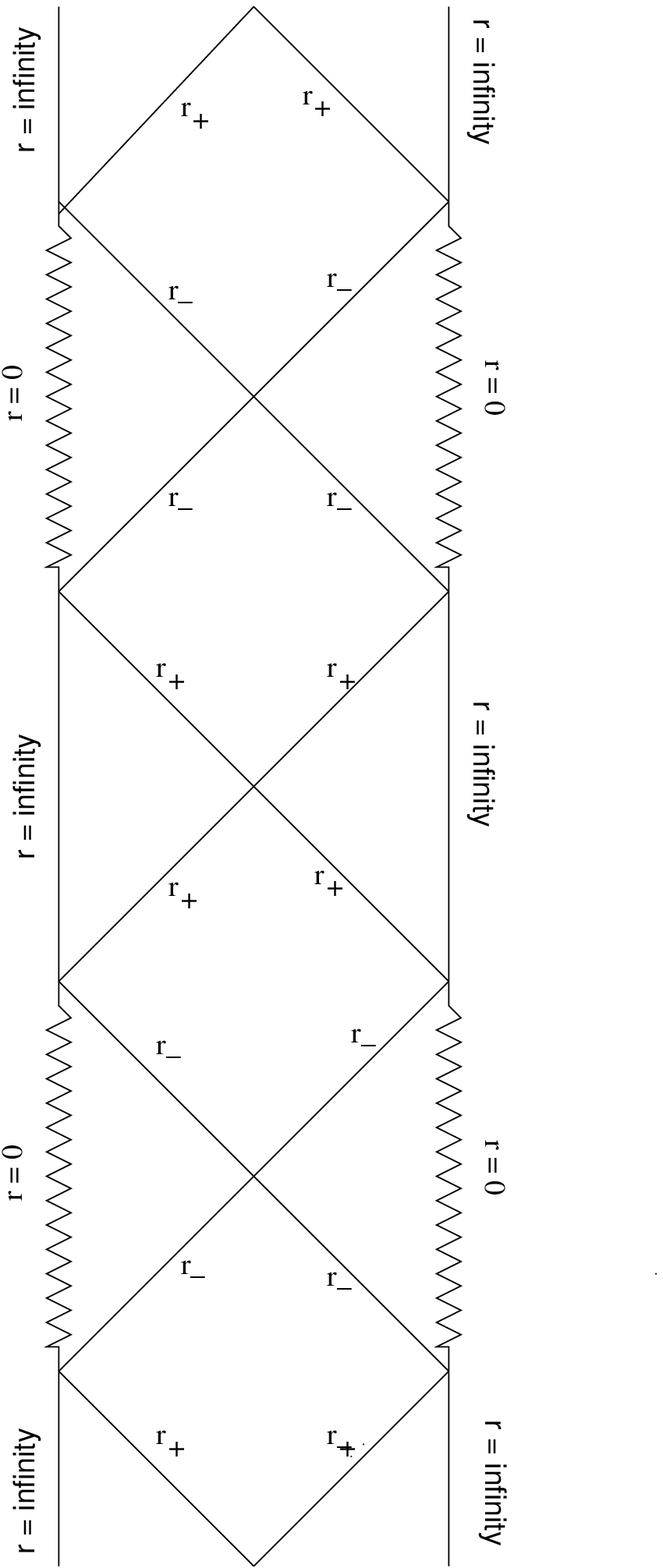}}
\caption{Penrose diagram for negative $\Lambda $ and $\protect\alpha <1$.}
\label{Fig1}
\end{figure}

\begin{figure}[tbp]
\epsfxsize=5cm \centerline{\epsffile{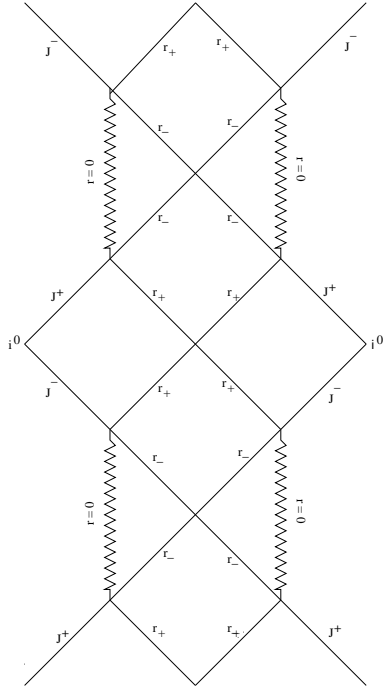}}
\caption{Penrose diagram for negative $\Lambda $ and $1\leq \protect\alpha <%
\protect\sqrt{n}$ .}
\label{Fig2}
\end{figure}

$\alpha ^{2}<n$: For $\alpha ^{2}<n$, as $r$ goes to infinity the dominant
term is the first term ($\Lambda $ term), and therefore there exist a
cosmological horizon for $\Lambda >0$, while there is no cosmological
horizons if $\Lambda <0$ . Indeed, in the latter case ($\alpha ^{2}<n$ and $%
\Lambda <0$) the spacetimes associated with the solution (\ref{Ur})-(\ref{Pr}%
) exhibit a variety of possible causal structures depending on the values of
the metric parameters $\alpha $, $m$, $q$ and $\Lambda $. One can obtain the
causal structure by finding the roots of $V(r)=0$. Unfortunately, because of
the nature of the exponents and hypergeometric function in (\ref{Vr}), it is
not possible to find explicitly the location of horizons. But, we can obtain
some information by considering the temperature of the horizons. Here, we
draw the Penrose diagram to show that the casual structure is asymptotically
well behaved. For reason of economy, we draw the Penrose diagram only for
the solution that presents a black brane with inner and outer horizons
(negative $\Lambda $ and $\alpha <\sqrt{n}$). The causal structure can be
constructed following the general prescriptions indicated in \cite{Walker}.
The Penrose diagram is shown in Figs. \ref{Fig1} and \ref{Fig2} for $\alpha
<1$ and $1\leq \alpha <\sqrt{n}$ respectively.
Also it is worth to write down the asymptotic behavior of the Ricci scalar.
Indeed, the form of the Ricci scalar for large values of $r$ is:
\begin{equation}
R=-\frac{n^{2}}{(n-3)^{2}l^{2}}\frac{(2\alpha ^{2}+n-3)[4\alpha
^{2}+(n+1)(n-3)]}{n-\alpha ^{2}}\left( \frac{c}{r}\right) ^{2(n-1)\gamma
/(n-3)}
\end{equation}%
which does not approach a nonzero constant as in the case of asymptotically
AdS spacetimes. It is worth to mention that the Ricci scalar of the solution
(\ref{Met3})-(\ref{Hr}) goes to zero as r goes to infinity, but with a
slower rate than that of an asymptotically flat spacetimes in the absence of
the scalar field.

Equation (\ref{Tem}) shows that the temperature is negative for the two
cases of (\emph{i}) $\alpha >\sqrt{n}$ despite the sign of $\Lambda $, and (%
\emph{ii}) positive $\Lambda $ despite the value of $\alpha $. As we argued
above in these two cases we encounter with cosmological horizons, and
therefore the cosmological horizons have negative temperature. Indeed, the
metric of Eqs. (\ref{Met3})-(\ref{Pr}) has two inner and outer horizons
located at $r_{-}$ and $r_{+}$, provided the mass parameter $m$ is greater
than $m_{\mathrm{ext}}$, an extreme black brane in the case of $m=m_{\mathrm{%
ext}}$, and a naked singularity if $m<m_{\mathrm{ext}}$ only for negative $%
\Lambda $ and $\alpha <\sqrt{n}$ where
\begin{eqnarray}
m_{\mathrm{ext}} &=&\frac{4(1+\alpha ^{2})^{2}}{r_{\mathrm{ext}}^{\gamma
(n-1)-n}}\left( \frac{\left( \frac{r_{\mathrm{ext}}}{c}\right) ^{-2\gamma
}\Lambda }{2(n-1)(\alpha ^{2}-n)}+\frac{(n-3)\beta ^{2}\left( \frac{r_{%
\mathrm{ext}}}{c}\right) ^{2\gamma (n+1)/(n-3)}(1-\sqrt{1+\eta _{\mathrm{ext}%
}})}{(n-1)\lambda }\right.  \nonumber \\
&&\left. +\frac{r_{\mathrm{ext}}^{2[\gamma (n-2)-(n-1)]}q^{2}\text{{\ }}%
_{2}F_{1}\left( \left[ \frac{1}{2},\frac{(n-3)\Upsilon }{2(n-1)}\right] ,%
\left[ 1+\frac{(n-3)\Upsilon }{2(n-1)}\right] ,-\eta _{\mathrm{ext}}\right)
}{\lambda c^{2\gamma (n-2)}\Upsilon }\right)  \label{mext}
\end{eqnarray}%
in Eq. (\ref{mext}), $r_{\mathrm{ext}}$ is the root of temperature relation (%
\ref{Tem}) such that
\begin{equation}
\left[ 1-\frac{\Lambda }{2\beta ^{2}}\left( \frac{r_{\mathrm{ext}}}{c}%
\right) ^{-8\gamma /(n-3)}\right] ^{2}-1=\eta _{\mathrm{ext}},
\end{equation}%
where
\begin{eqnarray*}
\eta _{\mathrm{ext}}=\frac{q^{2}\left( \frac{r_{\mathrm{ext}}}{c}\right)
^{2\gamma (n-1)(n-5)/(n-3)}}{\beta ^{2}r_{\mathrm{ext}}^{2(n-1)}}.
\end{eqnarray*}%
Note that in the absence of scalar field ($\alpha =\gamma =0$) $m_{\mathrm{%
ext}}$ reduces to that obtained in \cite{Deh3}.

Next, we calculate the electric charge of the solutions. According to the
Gauss theorem, the electric charge is the projections of the electromagnetic
field tensor on special hypersurfaces. Denoting the volume of the
hypersurface boundary at constant $t$ and $r$ by $V_{n-1}=(2\pi )^{k}\omega
_{n-k-1}$, the electric charge per unit volume $V_{n-1}$ can be found by
calculating the flux of the electric field at infinity, yielding%
\begin{equation}
{Q}=\frac{\Xi q}{4\pi l^{n-2}}
\end{equation}%
Comparing the above charge with the charge of black brane solutions of
Einstein-Born--Infeld-dilaton gravity, one finds that charge is invariant
under the conformal transformation (\ref{conf}). The electric potential $U$,
measured at infinity with respect to the horizon, is defined by \cite{Cal}
\begin{equation}
U=A_{\mu }\chi ^{\mu }\left\vert _{r\rightarrow \infty }-A_{\mu }\chi ^{\mu
}\right\vert _{r=r_{+}},  \label{Ch}
\end{equation}%
where $\chi $ is the null generators of the event horizon (\ref{Kil}). One
can easily show that the vector potential $A_{\mu }$ corresponding to
electromagnetic tensor (\ref{Ftr}) can be written as
\begin{equation}
A_{\mu }=\frac{qc^{(3-n)\gamma }}{\Gamma r^{\Gamma }}\text{{\ }}%
_{2}F_{1}\left( \left[ \frac{1}{2},\frac{(n-3)\Upsilon }{2(n-1)}\right] ,%
\left[ 1+\frac{(n-3)\Upsilon }{2(n-1)}\right] ,-\eta \right) \left( \Xi
\delta _{\mu }^{t}-a_{i}\delta _{\mu }^{i}\right) \hspace{0.5cm}{\text{(no
sum on i)}},  \label{Pot}
\end{equation}%
where $\Gamma =(n-3)(1-\gamma )+1$. Therefore the electric potential is
\begin{equation}
U=\frac{qc^{(3-n)\gamma }}{\Xi \Gamma {r_{+}}^{\Gamma }}\text{{\ }}%
_{2}F_{1}\left( \left[ \frac{1}{2},\frac{(n-3)\Upsilon }{2(n-1)}\right] ,%
\left[ 1+\frac{(n-3)\Upsilon }{2(n-1)}\right] ,-\eta _{+}\right)  \label{U}
\end{equation}

\section{Action and Conserved Quantities\label{Therm}}

The action (\ref{I1}) does not have a well-defined variational principle, we
should add the boundary action to it for ensuring well-defined
Euler-Lagrange equations. The suitable boundary action is
\begin{equation}
I_{b}=-\frac{1}{8\pi }\int_{\partial \mathcal{M}}d^{n}x\sqrt{-\gamma }K\Phi ,
\label{Ib1}
\end{equation}
where $\gamma $ and $K$ are the determinant of the induced metric and the
trace of extrinsic curvature of boundary. In general the action $I_{G}+I_{b}$
, is divergent when evaluated on the solutions, as is the Hamiltonian and
other associated conserved quantities. For asymptotically (A)dS solutions of
Einstein gravity, the way that one deals with these divergences is through
the use of counterterm method inspired by (A)dS/CFT correspondence \cite%
{Malda}. Although, in the presence of a non-trivial BD scalar field with
potential $V(\Phi )=2\Lambda \Phi ^{2}$, the spacetime may not behave as
either dS ($\Lambda >0$) or AdS ($\Lambda <0$). In fact, it has been shown
that with the exception of a pure cosmological constant potential, where $%
\alpha =0$, no AdS or dS static spherically symmetric solution exist for
Liouville-type potential \cite{Pollet}. But, as in the case of
asymptotically AdS spacetimes, according to the domain-wall/QFT (quantum
field theory) correspondence \cite{Boon}, there may be a suitable
counterterm for the action which removes the divergences. Since our
solutions have flat boundary [$R_{abcd}(h)=0$], there exists only one
boundary counterterm
\begin{equation}
I_{\mathrm{ct}}=-\frac{1}{8\pi }\int_{\partial \mathcal{M}}d^{n}x \sqrt{%
-\gamma }\frac{(n-1)}{l_{\mathrm{eff}}},  \label{Ict}
\end{equation}
where $l_{\mathrm{eff}}$ is given by
\begin{equation}
l_{\mathrm{eff}}^{2}=\frac{(n-1)(\alpha ^{2}-n)}{2\Lambda \Phi ^{3}}.
\label{Leff}
\end{equation}
One may note that as $\alpha $ goes to zero, the effective $l_{\mathrm{eff}%
}^{2}$ of Eq. (\ref{Leff}) reduces to $l^{2}=-n(n-1)/2\Lambda $ of the (A)dS
spacetimes. The total action, $I$, can be written as
\begin{equation}
I=I_{G}+I_{b}+I_{\mathrm{ct}}.  \label{Itot}
\end{equation}
The Euclidean actions (\ref{Itot}) per unit volume $V_{n-1}$ can be obtained
as
\begin{eqnarray}
I &=&\mathcal{\beta }_{+}\frac{c^{(n-1)\gamma }}{4\pi l^{n-2}}\Big[\frac{
(4\gamma +n-3)q^{2}{r_{+}^{-\Gamma }}}{\lambda \Gamma (\gamma -1)c^{2\gamma
(n-2)}}\text{{\ }}_{2}F_{1}\left( \left[ \frac{1}{2},\frac{(n-3)\Upsilon }{
2(n-1)}\right] ,\left[ 1+\frac{(n-3)\Upsilon }{2(n-1)}\right] ,-\eta
_{+}\right)  \nonumber \\
&&+\frac{(1+\alpha ^{2})r_{+}^{n-\gamma (n-1)}}{2(n-1)}\left( \frac{(\alpha
^{2}-1)}{(\alpha ^{2}-n)}\left[ \frac{r_{+}}{c}\right] ^{-2\gamma }\Lambda
-\right.  \nonumber \\
&&\left. \frac{2\left[ \lambda +(n-1)(n-3)\right] \beta ^{2}}{\lambda \left[
\frac{r_{+}}{c}\right] ^{-2\gamma (n+1)/(n-3)}}\left[ 1-\sqrt{1+\eta _{+}} %
\right] \right) \Big]  \label{finiteACT}
\end{eqnarray}
It is easy to show that the mass $M$ and the angular momentum $J_{i}$
calculated in Jordan (or string) frame is the same as Einstein frame and
they are remain unchanged under conformal transformations, that is
\begin{eqnarray*}
J_{i} &=&\frac{c^{(n-1)\gamma }}{16\pi l^{n-2}}\left( \frac{n-\alpha ^{2}}{
1+\alpha ^{2}}\right) \Xi ma_{i}, \\
M &=&\frac{c^{(n-1)\gamma }}{16\pi l^{n-2}}\left( \frac{(n-\alpha ^{2})\Xi
^{2}+\alpha ^{2}-1}{1+\alpha ^{2}}\right) m
\end{eqnarray*}
For $a_{i}=0$ ($\Xi =1$), the angular momentum per unit length vanishes, and
therefore $a_{i}$ is the $i$th rotational parameter of the spacetime.

We calculate the entropy through the use of Gibbs-Duhem relation
\begin{equation}
S=\frac{1}{T}(\mathcal{M}-\Gamma _{i}\mathcal{C}_{i})-I,  \label{GibsDuh}
\end{equation}
where $I$ is the finite total action (\ref{finiteACT}) evaluated on the
classical solution, and $C_{i}$ and $\Gamma _{i}$ are the conserved charges
and their associate chemical potentials respectively. It is straightforward
to show that
\begin{equation}
S=\frac{\Xi c^{(n-1)\gamma }}{4l^{\left( n-2\right) }}r_{+}^{(n-1)\left(
1-\gamma \right) },  \label{Entropy}
\end{equation}%
for the entropy per unit volume $V_{n-1}$. It is worth to note that the area
law is no longer valid in Brans-Dicke theory \cite{AreaLaw}. Nevertheless,
the entropy remains unchanged under conformal transformations.

Comparing the conserved and thermodynamic quantities calculated in this
section with those obtained in EBId gravity, one finds that they are
invariant under the conformal transformation (\ref{trans}). It is easy to
show that these quantities calculated satisfy the first law of
thermodynamics,
\begin{equation}
dM=TdS+{{{\sum_{i=1}^{k}}}}\Omega _{i}dJ_{i}+Ud{Q}  \label{First-law}
\end{equation}

\section{Closing Remarks}

The main goal of this paper is solving the field equations of BD theory in
the presence of non-linear electromagnetic field for an arbitrary value of $%
\omega $. As one can find, solving the field equations directly is a
non-trivial task because they include the second derivatives of the scalar
field. We could remove this difficulty through a conformal transformation.
Indeed, after conformal transformation, the BDBI action is reduced to EBId
action. We found that the suitable Lagrangian of EBId gravity is not the
same as the one presented in \cite{DHSR}, because it is not consistent with
conformal transformation. We also found analytical solutions of BDBI theory,
using conformal transformation (\ref{conf}) and investigated their
properties. Then we found that these solutions which exist only for $\alpha
^{2}\neq n$, have a cosmological horizon for $(i)$ $\alpha ^{2}>n$
despite(regardless of) the sign of $\Lambda $, and $(ii)$ positive values of
$\Lambda $, despite(disregarding) the magnitude of $\alpha $. For $\alpha
^{2}<n$, the solutions present black branes with inner and outer horizons if
$m>m_{ext}$, an extreme black brane if $m=m_{ext}$, and a naked singularity
otherwise. Also we presented Penrose diagrams and showed that the black
brane solutions are neither asymptotically flat nor (anti)-de Sitter. We
computed the finite Euclidean action through the use of counterterm method
and obtained the thermodynamic and conserved quantities of the solutions. We
found that the entropy does not follow the area law in BDBI theory but one
can show that it trace the area law in EBId gravity. One can find that the
conserved and thermodynamic quantities are invariant under the conformal
transformation and satisfy the first law of thermodynamics. The study of
spherical symmetric solutions of BDBI theory with non-zero curvature
boundary remains to be carried out in the future.

\begin{acknowledgements}
I am indebted to S. Bahaadini for her encouragements. This work
has been supported financially by Research Institute for Astronomy
and Astrophysics of Maragha.
\end{acknowledgements}

\end{document}